\documentclass[sigconf]{acmart}
\usepackage{booktabs} 

\usepackage[english]{babel}
\usepackage{moresize}
\usepackage{amsmath}
\usepackage[linesnumbered,algoruled,boxed,lined]{algorithm2e}
\usepackage{algorithmic}
\usepackage{balance}
\usepackage{comment}
\usepackage{paralist}
\usepackage{bm}
\usepackage{pgfplots}
\usetikzlibrary{pgfplots.dateplot}

\usepackage{flushend}
\usepackage[english]{babel}
\usepackage[latin1]{inputenc}
\usepackage{mathrsfs}
\usepackage{graphicx}

\usepackage{amssymb}
\usepackage{amsfonts}
\usepackage{url}
\usepackage{longtable}
\usepackage{rotating}
\usepackage{multirow}
\usepackage{mathrsfs}
\usepackage{subfigure}
\usepackage{enumitem}
\usepackage{adjustbox}
\usepackage{hyperref}
\usepackage{pgfplots}
\usetikzlibrary{pgfplots.dateplot}
\usepackage{filecontents}
\definecolor{tblue}{RGB}{31,119,180}
\definecolor{torange}{RGB}{255,127,14}
\definecolor{tgreen}{RGB}{44,160,44}
\definecolor{tred}{RGB}{214,39,40}
\definecolor{tpurple}{RGB}{148,103,189}

\newcommand{\hide}[1]{} 

\newcommand{\ie}{\textit{i}.\textit{e}.}
\newcommand{\eg}{\textit{e}.\textit{g}.} 
\newcommand{\wrt}{\textit{w}.\textit{r}.\textit{t}}

\def\model{MMSSL}

\copyrightyear{2023}
\acmYear{2023}
\setcopyright{acmlicensed}\acmConference[WWW'23]{Proceedings of the ACM Web Conference 2023}{May 1--5, 2023}{Austin, TX, USA}
\acmBooktitle{Proceedings of the ACM Web Conference 2023 (WWW'23), May 1--5, 2023, Austin, TX, USA}
\acmPrice{15.00}
\acmDOI{10.1145/3543507.3583206}
\acmISBN{978-1-4503-9416-1/23/04}

\begin{document}
\fancyhead{}



\title{Multi-Modal Self-Supervised Learning for Recommendation}



\author{Wei Wei}
\affiliation{%
  \institution{University of Hong Kong}
  \city{}
  \country{}}
\email{weiweics@connect.hku.hk}

\author{Chao Huang}
\authornote{Chao Huang is the corresponding author.}
\affiliation{%
  \institution{University of Hong Kong}
  \city{}
  \country{}}
\email{chaohuang75@gmail.com}

\author{Lianghao Xia}
\affiliation{%
  \institution{University of Hong Kong}
  \city{}
  \country{}}
\email{aka\_xia@foxmail.com}

\author{Chuxu Zhang}
\affiliation{%
  \institution{Brandeis University}
  \city{  }
  \country{  }}
\email{chuxuzhang@brandeis.edu}


\begin{abstract}
The online emergence of multi-modal sharing platforms (\eg, TikTok, Youtube) is powering personalized recommender systems to incorporate various modalities (\eg, visual, textual and acoustic) into the latent user representations. While existing works on multi-modal recommendation exploit multimedia content features in enhancing item embeddings, their model representation capability is limited by heavy label reliance and weak robustness on sparse user behavior data. Inspired by the recent progress of self-supervised learning in alleviating label scarcity issue, we explore deriving self-supervision signals with effectively learning of modality-aware user preference and cross-modal dependencies. To this end, we propose a new \underline{M}ulti-\underline{M}odal \underline{S}elf-\underline{S}upervised \underline{L}earning (\model) method which tackles two key challenges. Specifically, to characterize the inter-dependency between the user-item collaborative view and item multi-modal semantic view, we design a modality-aware interactive structure learning paradigm via adversarial perturbations for data augmentation. In addition, to capture the effects that user's modality-aware interaction pattern would interweave with each other, a cross-modal contrastive learning approach is introduced to jointly preserve the inter-modal semantic commonality and user preference diversity. Experiments on real-world datasets verify the superiority of our method in offering great potential for multimedia recommendation over various state-of-the-art baselines. The  implementation is released at: \url{https://github.com/HKUDS/MMSSL}.
\end{abstract}




\maketitle

\section{Introduction}
\label{sec:intro}

Multimedia recommender systems play a crucial role in a wide range of content-sharing and e-commerce applications with massive web multimedia content, including micro-videos, images and songs~\cite{qiu2021causalrec}. In multimedia recommendation scenarios, various modalities of item content are involved, such as the visual, acoustic, and textual features of items~\cite{wang2020online}. Such multi-modal data characteristics may reflect users' preferences with fine-grained modality level.

Several research lines have emerged to incorporate multi-modal content into multimedia recommendation. For example, VBPR~\cite{he2016vbpr} as an early study to extend matrix decomposition framework to deal with the modality features of items. ACF~\cite{chen2017attentive} proposes to identify the component-level user preference via a hierarchically structured attention network. To explore high-order connectivity among user-item interactions, recently proposed methods (\eg, MMGCN~\cite{wei2019mmgcn}, GRCN~\cite{wei2020graph}, LATTICE~\cite{zhang2021mining}) adopt Graph Neural Networks (GNNs) to incorporate modality information into the message passing for inferring user and item representations. However, the satisfied performance of most existing multimedia recommenders usually requires sufficient high-quality label data (\ie, observed user interactions) to train the models in a supervised manner. In real-life recommendation scenarios, interaction labels between users and items are very sparse compared with the whole interaction space~\cite{wu2021self,lin2022improving}, which limits the representation ability of current fully supervised models to generate accurate embeddings to represent complex user preference in multimedia recommendation.

Inspired by the recent success of self-supervised learning (SSL) for data augmentation~\cite{khosla2020supervised,zhu2021graph}, an antidote to mitigate the data sparsity limitation in multimedia recommendation is to generate supervisory signals from the unlabeled data. While some recent studies (\eg, SGL~\cite{wu2021self}, NCL~\cite{lin2022improving}, HCCF~\cite{xia2022hypergraph}) attempt to incorporate SSL into the modeling of user-item interactions for collaborative filtering, they fail to adapt the augmentation schemes to the specific multimedia recommendation task. For example, SGL~\cite{wu2021self} directly performs stochastic noise perturbation to dropout nodes and edges for graph augmentation. NCL~\cite{lin2022improving} and HCCF~\cite{xia2022hypergraph} propose to discover implicit semantic node correlations over the observed user-item interactions. The overlook of multi-modal characteristics for augmentation may hinder the effectiveness of their introduced auxiliary SSL tasks to distill modality-aware signals.

\begin{figure}[t]
	\centering
    \includegraphics[width=0.9598\columnwidth]{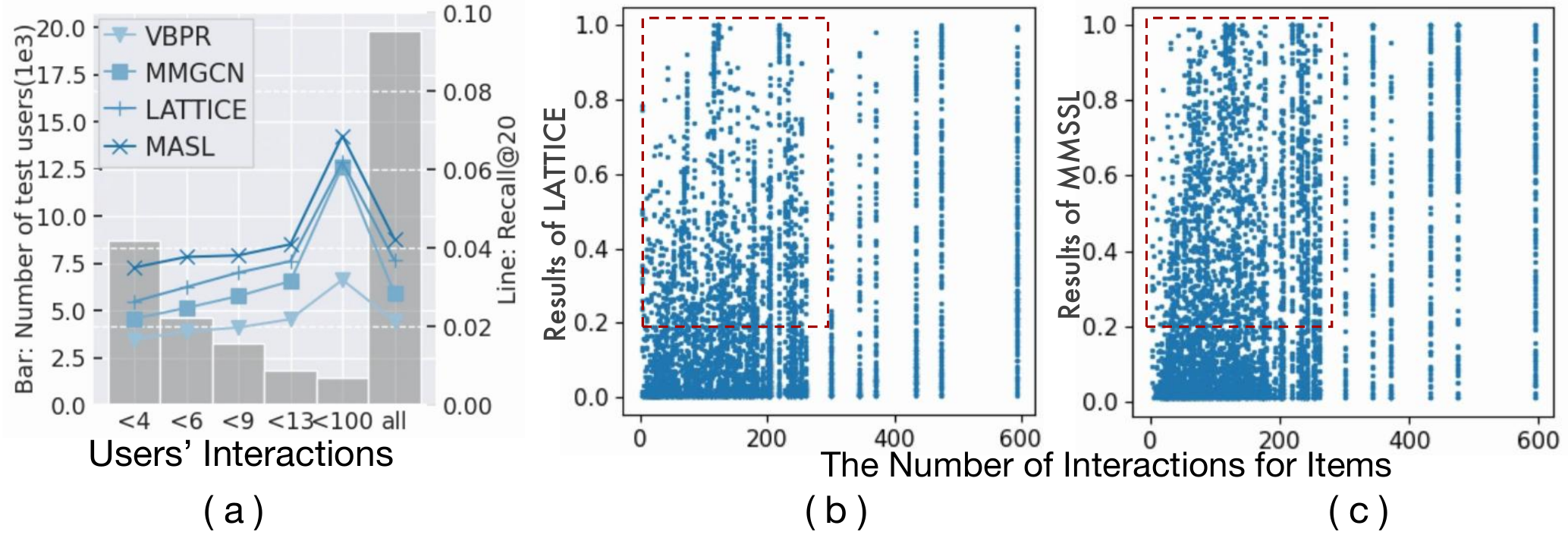}
	\vspace{-0.2 in}
	\caption{(a) Performance \wrt\ user interaction frequencies. Left side y-axis: \# of users in each group; Right side y-axis: performance of different methods. (b)-(c) Performance \wrt\ item sparsity degrees. Each point represents item-specific recommendation accuracy averaged across epochs. 
 }
    \Description[Performance \wrt\ user interaction frequencies]{Left side y-axis: \# of users in each group; Right side y-axis: performance of different methods. (b)-(c) Performance \wrt\ item sparsity degrees. Each point represents item-specific recommendation accuracy averaged across epochs. Baselines perform worse on sparse data.}
	\label{fig:sparsity}
	\vspace{-0.25in}
     \vspace{-0.1em}
\end{figure}

Consider the presented the results on Amazon-baby data in Figure~\ref{fig:sparsity}, several state-of-the-art multimedia methods are experimentally compared to make recommendation with respect to different sparsity degrees from user and item side. To be specific, we notice that user interaction sparsity hinders the representation ability of compared methods to capture the genuine multi-modal preference of users. Significant performance gain is achieved by our \model\ with highly sparse user interaction data (\eg, $<4$). By visualizing the performance distribution of item-specific prediction results, our method can be observed with more long-tail items distributed in the highlighted area. This gives an illustrated understanding of our model superiority in alleviating long-tail issue for recommendation. \\\vspace{-0.12in}


To address the above limitations, this work explores multi-modal self-supervised learning paradigm to benefit multimedia recommendation from two perspectives of modality-aware augmentation.\\\vspace{-0.12in}

\noindent \textbf{Modality-aware Collaborative Relation Learning}. In multimedia recommendation, the diversity of modalities can reflect different user preferences over items, which could be the interests in visual features of videos, acoustic characteristics of songs, and textual description of products~\cite{wang2021dualgnn}. Failing to inject modality-aware collaborative signals into the self-supervised learning task is insufficient to distill effective SSL signals. Hence, to perform the data augmentation with the awareness of modality-aware user preference, we propose an adversarial self-supervised learning method to explore the implicit relationships between users and items at the fine-grained modality-specific level. To distill the self-supervision signals pertinent to modality-aware user-item interaction patterns, our generative SSL-based augmentor integrates the modality-guided collaborative relation generator and discriminator, so as to model the influence of multi-modal content on user interaction behaviors. \\\vspace{-0.12in} 

\noindent \textbf{Cross-Modality Dependency Modeling}. Given that different modality-specific user preferences would interweave with each other in an implicit manner~\cite{truong2021multi}, leaving this fact untouched can easily lead to less accurate representations in preserving multi-modal user-item relationships. For instance, users may get attracted by a micro-video due to both its amazing visual content and instrumental background music. Additionally, purchase behaviors of customers in online retailers could be influenced by the presented product images as well as the posted product reviews. To enable our auxiliary supervision signals to be reflective of the modality-wise pattern influence, we propose to enhance our self-supervised learning paradigm with cross-modal dependency modeling.\\\vspace{-0.12in}

In light of the aforementioned motivations for model design, we develop \model, a new multimedia recommender that unifies the generative and contrastive SSL for modality-aware augmentation. We pursue a generic solution to jointly capture modality-specific collaborative effects and cross-modality interaction dependency, in recognizing multi-modal user preference. In particular, at the first stage of our SSL paradigm, we propose to train an adversarial relation learning network with the perturbed modality-aware user-item dependency weights. This scheme allows us to distill the useful multi-modal information and encode them into the latent user (item) embeddings facing the limitation of label scarcity. To tackle the challenge of adversarial learning over a sparse user-item connection matrix, we integrate the Gumbel-based projection and Wasserstein adversarial generation to mitigate the distribution gap. At the second stage of \model, we introduce a cross-view multi-modal contrastive learning scheme to pursue encoding the influence among modality-wise preference into representations with model robustness. Furthermore, we offer the theoretical discussion of our self-supervised learning paradigm from viewpoints of: i) enhancing the multi-modal knowledge transfer to the collaborative view via adversarial self-augmentation; ii) benefiting the gradient learning with the cross-modal contrastive augmentation. 

In a nutshell, our contributions can be summarized as follows.
\begin{itemize}[leftmargin=*]
\item We tackle the label scarcity issue of multimedia recommendation with dual-stage self-supervised learning for modality-aware data augmentation. By unifying the recommendation task and our augmented generative/contrastive multi-modal SSL signals, significant performance gains can be achieved by our \model.\\\vspace{-0.12in}
\item We propose a new recommender system \model\ which integrates the generative modality-aware collaborative self-augmentation and the contrastive cross-modality dependency encoding. \\\vspace{-0.12in}
\item We extensively evaluate the proposed \model\ to justify the model's effectiveness and robustness. In-depth and visual analyses demonstrate the rationality of our \model\ method.
\end{itemize}


\vspace{-0.8 em}

\section{Preliminary}
\label{sec:model}

\noindent \textbf{Interaction Graph with Multi-Modality}. In the context of graph learning, GNN-based collaborative filtering paradigms have offered state-of-the-art results~\cite{he2020lightgcn,wang2019neural}. Inspired by this, \model\ is built over the graph-structured interaction data. Specifically, we generate an user-item graph $G=\{(u,i)|u\in \mathcal{U}, i\in \mathcal{I}\}$, where $\mathcal{U}$ and $\mathcal{I}$ represents the set of users and items, respectively. We construct edge in $G$ between user $u$ and item $i$ if an interaction is observed. We incorporate the multi-modal information (\eg, textual, visual, acoustic modality) into the generated user-item interaction graph $G$ with modality-aware characteristic features $\bar{\textbf{F}} = \{\bar{\textbf{f}}_i^1, ..., \bar{\textbf{f}}_i^m ,..., \bar{\textbf{f}}_i^{ |\mathcal{M} | }\}$ of item $i$. Here, $\bar{\textbf{f}}_i^m \in\mathbb{R}^{d_m} $ represents the raw feature embedding (with dimensionality $d_m$) of item $i$ with modality $m\in \mathcal{M}$ .\\\vspace{-0.12in}

\noindent \textbf{Task Formulation}. We formulate our multi-modal recommender system that captures user-item relations with modality-aware user preference learning. In particular, given the generated multi-modal interaction graph $G=\{(u,i)|u\in \mathcal{U}, i\in \mathcal{I}, \bar{\textbf{F}}\}$, our task of multi-modal recommender is to learn a function that forecasts how likely an item will be adopted by an user.

\section{Methodology}
\label{sec:solution}

\begin{figure*}
    \centering
    \includegraphics[width=\textwidth]{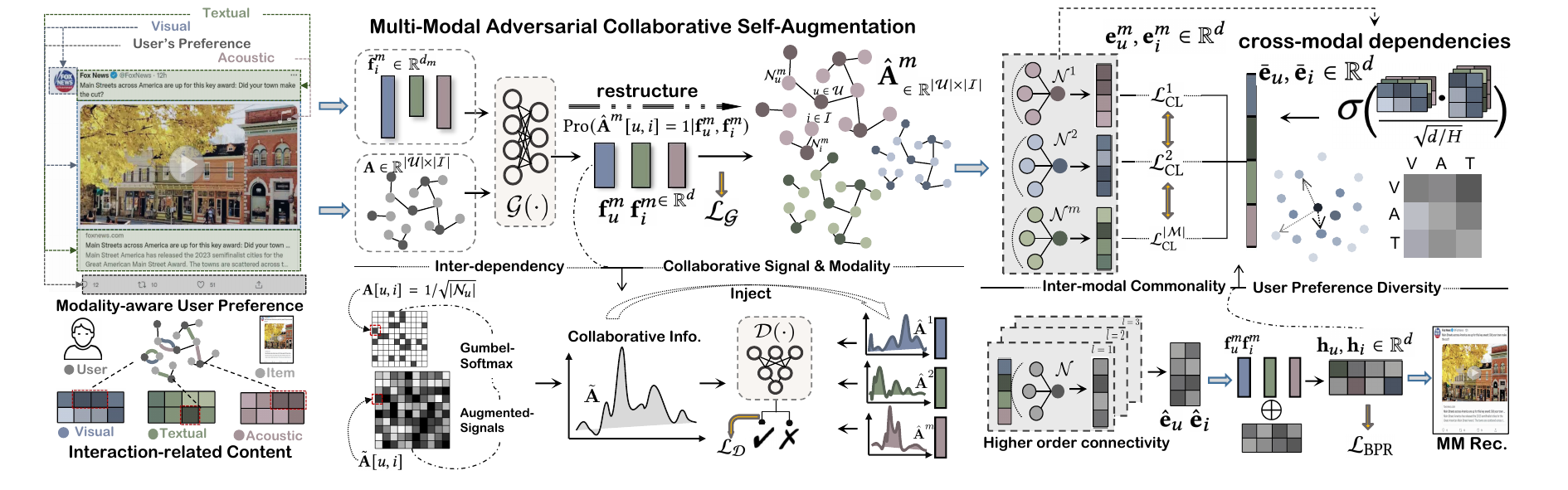}
    \vspace{-0.35in}
    \caption{The model flow of our \model. Gumbel-based transformation is integrated with Wasserstein adversarial generation to mitigate distribution gap between our augmented user-item relation matrix $\hat{\textbf{A}}^m$ generated via $\mathcal{G}(\cdot)$ and the original $\textbf{A}$.
    }
    \Description[Performance \wrt\ user interaction frequencies]{Gumbel-based transformation is integrated with Wasserstein adversarial generation to mitigate distribution gap between our augmented user-item relation matrix $\hat{\textbf{A}}^m$ generated via $\mathcal{G}(\cdot)$ and the original $\textbf{A}$.}
    \vspace{-0.9 em}
    \label{fig:framework}
\end{figure*}

\subsection{Multi-Modal Self-Augmentation}
\label{sec:ad-module}
In our model, we design a self-supervised learning task to supplement the user-item interaction modeling with multi-modal adversarial user-item relation learning. Benefiting from such design, our \model\ not only captures the modality-aware user preference over items but also effectively alleviates the data scarcity with the derived self-supervision signals from multi-modal context. To achieve our goal, we propose an adversarial self-augmentation method which is composed of the modality-guided collaborative relation generator $\mathcal{G}(\cdot)$ and graph structure discriminator $\mathcal{D}(\cdot)$.

\subsubsection{\bf Modality-guided Collaborative Relation Generator}
\label{sec:cl-theory}
In the generative stage of our self-augmentation scheme, we aim to perform modality-aware graph structure learning over user-item interactions, so as to excavate modality-specific user preferences. In other words, we aim to derive the likelihood of user $u$ interacting with item $i$ given the corresponding multi-modal context:
\begin{align}
\text{Pro}(\hat{\textbf{A}}^m \vert \textbf{F}^m)  =  \text{Pro}(\hat{\textbf{A}}^m[u,i] = 1 \vert \textbf{f}_u^m, \textbf{f}_i^m)
\end{align}
\noindent where $\hat{\textbf{A}}^m\in\mathbb{R}^{|\mathcal{U}|\times |\mathcal{I}|}$ represents the learned user-item interaction matrix under the modality $m$. Each element in $\hat{\textbf{A}}^m$ is denoted as $\hat{\textbf{A}}^m[u,i]\in\mathbb{R}$. To generate the input modality-aware user and item representations $\textbf{f}_u^m$ and $\textbf{f}_i^m$, we incorporate the collaborative effects into embeddings based on raw multi-modal feature vector $\bar{\textbf{f}}_i^m$:
\begin{align}
\label{eq:quality-prior}
    \textbf{f}_u^m = \sum_{i\in\mathcal{N}_u} \bar{\textbf{f}}_i^m / \sqrt{|\mathcal{N}_u|};~~~~~~
    \textbf{f}_i^m = \sum_{u\in\mathcal{N}_i} \textbf{f}_u^m / {\sqrt{|\mathcal{N}_i|}}
\end{align}
\noindent Here, $\mathcal{N}_u$ and $\mathcal{N}_i$ denotes the neighborhood set of user $u \in \mathcal{U} $ and item $i \in \mathcal{I}$ in the user-item interaction graph $G$. Before feeding the multi-modal feature vectors into our generator, we use the fully-connected layer with dropout~\cite{hinton2012improving} to perform a modality-specific transformation to map raw multi-modal features into latent embedding space, \ie, $\bar{\textbf{f}}_i^m \in\mathbb{R}^{d_m} \longrightarrow \bar{\textbf{f}}_i^m \in\mathbb{R}^{d}, d \ll d_m$. 

To capture the inter-dependency between the collaborative relationships and multi-modal contextual features, the modality-aware graph structure learning is conducted by learning the preference matrix $\hat{\textbf{A}}^m\in\mathbb{R}^{|\mathcal{U}|\times |\mathcal{I}|}$ with multi-modal representations:
\begin{align}
\label{eq:fake-data}
    \hat{\textbf{A}}^m[u,i] = \mathcal{G}(\textbf{f}_u^m, \textbf{f}_i^m) = {\textbf{f}_u^m}^\top \cdot {\textbf{f}_i^m} / (\|\textbf{f}_u^m\|_2 \cdot \|\textbf{f}_i^m\|_2)
\end{align}
\noindent To avoid calculating the entire interaction matrix, we conduct batch-based block matrix multiplication for memory-efficient implementation. With our collaborative relation generator $\mathcal{G}(\cdot)$, the learned modality-specific user-item relation matrix $\hat{\textbf{A}}^m$ captures the inter-dependency between the collaborative view and multi-modal context view, which is reflective of modality-aware user preference.

\subsubsection{\bf Discriminator with Adversarial Generation}
In our adversarial SSL paradigm, the discriminator $\mathcal{D}(\cdot)$ is designed to distinguish the generated modality-aware user-item relations $\hat{\textbf{A}}^m$ and the observed user-item interaction matrix $\textbf{A}$ from graph $G$. Through our relation discrimination process, our generator tries to confuse the discriminator by refining the learned modality-aware relation matrix. By doing so, the implicit inter-dependency modeling between the collaborative view and multi-modal context view can be enhanced by achieving adversarial robustness in representations. Formally, the discriminator $\mathcal{D}(\cdot)$ is built as follows:
\begin{equation}
\label{eq:d-model}
    \mathcal{D}(\textbf{a}) = \delta(\Gamma^2(\textbf{a}));~
    \Gamma(\textbf{a}) = \text{Drop}(\text{BN}(\text{LeakyRelu}( \text{Linear}(\textbf{a})   )))
\end{equation}
\noindent Following the learning strategy in~\cite{metz2016unrolled,radford2015unsupervised}, the input embedding $\textbf{a} \in\mathbb{R}^{|\mathcal{I}|} $ corresponds to each matrix row, which is selected from either the generated relations $\hat{\textbf{A}}^m$ or observed interactions $\textbf{A}$, \ie, $ \{ \textbf{a} \ | \textbf{a} \in \textbf{A} \ or  \  \textbf{a} \in \hat{\textbf{A}}^m ,m\in { \mathcal{M} } \} $. $\Gamma(\cdot)$ denotes the discrimination neural layer by utilizing i) batch-normalization $\text{BN}(\cdot)$ for preventing the gradient vanishing issue~\cite{ioffe2015batch}; ii) dropout $Drop(\cdot)$ for alleviating overfitting~\cite{hinton2012improving}; iii) non-linear activation ${LeakyRelu}(\cdot)$ for facilitating model convergence with more complete gradients~\cite{maas2013rectifier,hu2018generative}. We stack two fully connected layers in our discriminator and adopt the sigmoid function $\delta(\cdot)$ to approximate the distribution of binary interaction data. 
${Linear}(\cdot)$ applies a linear transformation to $\textbf{a}$.

\subsubsection{\bf Adversarial SSL against Distribution Gap}
\label{sec:ad-gap}
Different from the dense pixel matrix of vision data, the observed user-item interaction matrix is highly sparse by including large number of zero values. Such sparsity poses unique challenges for our adversarial modality-aware relation learning, due to the data distribution difference between our generator $\mathcal{G}(\cdot)$ and discriminator $\mathcal{D}(\cdot)$. In particular, the user-item relation matrix $\hat{\textbf{A}}^m$ learned by the generator with dense values is inevitably quite different from the observed interaction matrix $\textbf{A}$, which may lead to the mode collapse and difficulty with convergence~\cite{srivastava2017veegan}.

To address this challenge, we leverage \emph{Gumbel-Softmax}~\cite{jang2016categorical} to transform the original interaction data into a dense matrix based on the Gumbel distribution, and bridge the distribution gap limitation. To be specific, the adversarial enhancement is presented:
\begin{equation}
\label{eq:real-data}
    \begin{aligned}
    \tilde{\textbf{A}}[u,i] &= \begin{matrix} \underbrace{ \frac{ \exp{\left( \left( \textbf{A}[u,i]+g \right)/\tau \right)} }{ \sum_{i'}{\exp{\left( \left( \textbf{A}[u,i']+g\right)/\tau \right)}} }  } \\  \ transformation \end{matrix}  +   \begin{matrix} \underbrace{  \zeta {\times}{ \frac{\textbf{h}_u^\top \textbf{h}_i}{\|\textbf{h}_u\|_2\|\textbf{h}_i\|_2}}} \\ augmented\ signals\ \end{matrix} 
    \end{aligned}
\end{equation}
\noindent Here, the perturbation factor $g$ is calculated as $g =- log(-log(u))$ based on Gumbel (0, 1) distribution, where u $\sim$ Uniform(0, 1)~\cite{jang2016categorical}. With the Gumbel-based transformation, the original observed interaction matrix $\textbf{A}$ can be projected into a closely distributed matrix $\tilde{\textbf{A}}\in\mathbb{R}^{|\mathcal{U}|\times|\mathcal{I}|}$. $\tau$ is the temperature factor to adjust the smoothness. To further improve the robustness of our adversarial SSL, we inject the auxiliary signals with the final multi-modal collaborative embeddings $\textbf{h}_u, \textbf{h}_i$ derived from Eq.~\ref{eq:final-embed}. $\zeta$ is a weight parameter.

\subsubsection{\bf Adversarial SSL Loss}
In our adversarial SSL task, we aim to capture the inter-dependency between the collaborative view and multi-modal view by aligning the distributions between our learned modality-aware user-item relations $\hat{\textbf{A}}^m$ and the proxy $\tilde{\textbf{A}}$ of the observed user-item interactions $\textbf{A}$. Towards this end, we define our adversarial SSL loss to optimize our relation generator $\mathcal{G}(\cdot)$ and discriminator $\mathcal{D}(\cdot)$ in a minimax optimization manner as follows:
\begin{equation}
    \begin{split}
    \min\limits_{\mathcal{G}} \max\limits_{\mathcal{D}}\mathbb{E}_{\tilde{\textbf{A}}\sim \mathbb{P}_r}[ \mathcal{D}(\tilde{\textbf{A}})] - \mathbb{E}_{\hat{\textbf{A}}^m \sim \mathbb{P}_f}[(\mathcal{D}( \mathcal{G}(\textbf{f}^m) )] 
    \end{split}
\end{equation}
\noindent we separately present the optimized objectives $\mathcal{L}_\text{G}$ and $\mathcal{L}_\text{D}$ corresponding to the generator $\mathcal{G}$ and discriminator $\mathcal{D}$, respectively.
\begin{small}
\begin{equation}
 \label{eq:gan-loss}
    \begin{cases}
    \mathcal{L}_\mathcal{G}=-\mathbb{E}_{\hat{\textbf{A}}^m}[\mathcal{D}(\hat{\textbf{A}}^m)]\\
    \mathcal{L}_\mathcal{D} = - \mathbb{E}_{\tilde{\textbf{A}}}[\mathcal{D}(\tilde{\textbf{A}})] + \mathbb{E}_{\hat{\textbf{A}}^{m}}[\mathcal{D}(\hat{\textbf{A}}^m)] + \lambda_1  \mathbb{E}_{\ddot{\textbf{A}}}[(\|\nabla_{ \mathcal{D}(\ddot{\textbf{A}})} \| - 1)^2]
    \end{cases}
\end{equation}
\end{small}
\noindent To further enhance the robustness of our adversarial self-supervised learning against distribution gap and data sparsity, WassersteinGAN-GP~\cite{gulrajani2017improved} is introduced to add the gradient penalty with the balance weight $\lambda_1$. Here, $\ddot{\textbf{A}}$ denotes the interpolation of $\hat{\textbf{A}}^m$ and $\tilde{\textbf{A}}$ matrix.

\subsection{Cross-Modal Contrastive Learning}
In multimedia recommendation scenarios, user interaction behavioral patterns with different item modalities (\eg, visual, textual and acoustic) will influence each other. For example, the visual and acoustic features of a short video can jointly attract users to view it. Thus, the visual-specific and acoustic-specific preferences of users may interweave in a complex way. To capture such implicit dependencies among user modality-specific preferences, we design a cross-modal contrastive learning paradigm with modality-aware graph contrastive augmentation.

\subsubsection{\bf Modality-aware Contrastive View}
To inject the modality-specific semantics into our contrastive learning component, we perform the information aggregation over the modality-aware semantic neighbors $\mathcal{N}_u^m$ and $\mathcal{N}_i^m$ of user $u$ and item $i$. It can be derived from relational matrix $\hat{\textbf{A}}^m$ (Eq. \ref{eq:fake-data}) learned in our generator.
\begin{align}
    \label{eq:col_gcn}
    \textbf{e}_u^m=\sum_{i\in\mathcal{N}_u^m} \textbf{e}_i / \sqrt{|\mathcal{N}_u^m|};~~~~~~
    \textbf{e}_i^m=\sum_{u\in\mathcal{N}_i^m} \textbf{e}_u / \sqrt{|\mathcal{N}_i^m|}
\end{align}
\noindent $\textbf{e}_u , \textbf{e}_i \in \mathbb{R}^d $ are Xavier-initialized~\cite{glorot2010understanding} id-corresponding embeddings. The multi-modal contextual information can be preserved in the modality-aware latent representations $\textbf{e}_u^m, \textbf{e}_i^m \in \mathbb{R}^d $.\\\vspace{-0.12in}

\noindent \textbf{Modality-wise Dependency Modeling}. To capture the correlations between each pair of modality-specific user preferences, we design our modality-wise dependency encoder with a multi-head self-attention mechanism using the following formula:
\begin{align}
\label{eq:inter-modality}
    \bar{\textbf{e}}_u^m = \sum_{m'\in \mathcal{M}}^{|\mathcal{M}|} \mathop{\Bigm|\Bigm|}\limits_{h=1}^H \sigma( \frac{(\textbf{e}_u^m\textbf{W}_h^Q)^\top\cdot(\textbf{e}_u^{m'}\textbf{W}_h^K)}{\sqrt{d/H}})\cdot \textbf{e}_u^{m'}
\end{align}
\noindent where $\textbf{W}_h^Q, \textbf{W}_h^K\in\mathbb{R}^{d/H\times d}$ denote the $h$-head query and the key transformations for calculating the relation between modality pair $(m, m')$. $H$ denotes the number of attention heads. $\sigma(\cdot)$ denotes the softmax function. Embeddings of both user and item side are refined in an analogous way. We then fuse the modality-specific embeddings through mean-pooling (\eg~$\bar{\textbf{e}}_u=\sum_{m \in \mathcal{M}}^{|\mathcal{M}|}\bar{\textbf{e}}_u^m / {|\mathcal{M}|}$), to generate the multi-modal user/item representations $\bar{\textbf{e}}_u, \bar{\textbf{e}}_i\in\mathbb{R}^d$.\\\vspace{-0.12in}

\noindent \textbf{Multi-Modal High-Order Connectivity}. To explore the high-order collaborative effects with the awareness of multi-modal information, we build our encoder upon the graph neural network for recursive message passing with the matrix form as follows:
\begin{align} 
\label{eq:high-order}
    \hat{\textbf{E}}_u^{l+1} = \textbf{A} \cdot \hat{\textbf{E}}_i^l;~~~~~ \hat{\textbf{E}}_u^0=\textbf{E}_u + \eta \cdot \bar{\textbf{E}}_u / {\|\bar{\textbf{E}}_u\|_2^2}
\end{align} 
\noindent where $\hat{\textbf{e}}_i^{l} \in \hat{\textbf{E}}_i^l, \hat{\textbf{e}}_u^{l+1} \in \hat{\textbf{E}}_u^{l+1}$ denote the embeddings for the $l$-th and the $(l+1)$-th layer, respectively. The node degree-normalized matrix element $\textbf{A}[u,i]=1/\sqrt{|\mathcal{N}_u|}$ if user $u$ has interacted with item $i$ from the observed data. The zero-order embeddings $\hat{\textbf{E}}_u^0$ are obtained by combining the initial id-corresponding embeddings $\textbf{E}_u$ with the normalized modality-aware embeddings $\bar{\textbf{E}}_u$ using the weight parameter $\eta$. In our multi-layer GNNs, the layer-specific embeddings are aggregated through mean-pooling to yield the output: $\hat{\textbf{E}}_u = \sum_{l=0}^L \hat{\textbf{E}}_u^l / L$, where $L$ is the number of graph layers.

\subsubsection{\bf Cross-Modal Contrastive Augmentation}
In this module, we introduce our multi-modal contrastive learning paradigm which distills the self-supervision signals for dependency modeling among modality-specific user preferences. Specifically, we adopt the InfoNCE loss function to maximize the mutual information between the modality-specific embeddings $\textbf{e}_u^m$ and the overall user embedding $\textbf{h}_u$ of the same user $u$. With the self-discrimination strategy~\cite{wu2021self,xia2022hypergraph}, embeddings from different users are treated as negative pairs. Our cross-modal contrastive loss is defined as: 
\begin{align}
    \label{eq:infonce-loss}
    \mathcal{L}_\text{CL} = - & \sum_{m \in \mathcal{M}}^{|\mathcal{M}|} \sum_{u\in\mathcal{U}}^{|\mathcal{U}|} \log \frac{\exp s(\textbf{h}_u, \textbf{e}_u^m)}{\sum\limits_{u'\in\mathcal{U}} \left(\exp s(\textbf{h}_{u'}, \textbf{e}_u^m) + \exp s(\textbf{e}_{u'}^m, \textbf{e}_u^m) \right)}\nonumber\\
    &s(\textbf{h}_u, \textbf{e}_u^m)=\textbf{h}_u^\top\cdot\textbf{e}_u^m / (\tau'\cdot\|\textbf{h}_u\|_2\cdot\|\textbf{e}_u^m\|_2)
\end{align}
\noindent where $s(\cdot)$ denotes the similarity function, and $\tau'$ is the temperature coefficient. Our cross-modal contrastive learning aims to learn an embedding space in which different user representations are far apart, which allows our model to capture diverse user modality-specific preferences with uniformly-distributed embeddings. 

\subsection{Multi-Task Model Training}
To generate our final user (item) representations $\textbf{h}_u, \textbf{h}_i \in \mathbb{R}^d$ for making predictions, we promote the cooperation between the collaborative view and multi-modal view by aggregating their corresponding encoded embeddings as follows: 
\begin{align}
\label{eq:final-embed}
    \textbf{h}_u = \hat{\textbf{e}}_{u} + \omega \sum_{m \in \mathcal{M}}^{|\mathcal{M}|}{   \frac{\textbf{f}_{u}^{m}}{\|\textbf{f}_u^m\|_2}}; ~~~~
    \textbf{h}_i = \hat{\textbf{e}}_{i} + \omega \sum_{m \in \mathcal{M}}^{|\mathcal{M}|}{  \frac{\textbf{f}_{i}^{m}}{\|\textbf{f}_i^m\|_2}}
\end{align}
\noindent $\omega$ is the aggregation weight. Normalization is performed to alleviate the value scale difference between the embeddings of collaborative view ($\hat{\textbf{e}}_u, \hat{\textbf{e}}_i$) and multi-modal views ($\textbf{f}^m_u, \textbf{f}^m_i$). With the final embeddings, our \model\ model makes predictions on the unobserved interaction between user $u$ and item $i$ through $\hat{y}_{u,i}=\textbf{h}_u^\top \cdot \textbf{h}_i$.\\\vspace{-0.12in}

\noindent \textbf{Model Optimization}. We train our recommender with a multi-task learning scheme to jointly optimize \model\ with i) the main user-item interaction prediction task $\mathcal{L}_\text{BPR}$; ii) adversarial modality-aware user-item relation learning task $\mathcal{L}_\text{G}$; iii) cross-modal contrastive learning task $\mathcal{L}_\text{CL}$. Formally, the jointly optimized objective is given ($\lambda_2, \lambda_3, \lambda_4$ are hyperparameters for loss term weighting):
\begin{align}
\label{eq:total-loss}
\mathcal{L} = \mathcal{L}_\text{BPR} + \lambda_2\cdot \mathcal{L}_\text{CL} + \lambda_3\cdot \mathcal{L}_\text{G} + \lambda_4\cdot \|\Theta\|^2 \\
\mathcal{L}_\text{BPR} = \sum_{(u,i_p,i_n)}^{|\mathcal{E}|} - \log \left(\text{sigm} (\hat{y}_{u,i_p} - \hat{y}_{u,i_n})\right)
\end{align}
\noindent $i_p, i_n$ denotes the positive and negative samples for user $u$. The last term in $\mathcal{L}$ is the weight-decay regularization against over-fitting.

\subsection{Theoretical Discussion of \model}
We offer theoretical analyzes to discuss the benefits of our multi-modal SSL:
i) conducting collaborative knowledge transfer; ii) capturing the cross-modality commonality for each instance.

\subsubsection{\bf Theoretical Analysis of Knowledge Transfer}
The purpose of the adversarial SSL task is to inject collaborative signals into the modality-specific distribution (\ie, knowledge transfer). Theoretical analysis here supports for the associativity of adversarial transferability and knowledge transferability. 
In particular, we first provide formal definitions of those two transferability and then show their correlations. 
For knowledge transferability, we care about whether the source $\hat{\textbf{A}}^m$ on data $G$ can achieve low loss $\mathcal{L}(.; G )$ when combining with a trainable 
model $\mathcal{G}$ comparing with target $\tilde{\textbf{A}}$, which is formally shown as follows:
\begin{equation}
    \begin{split}
        \min_{ \mathcal{G} } \mathcal{L} ( \mathbf{G} \circ \hat{\textbf{A}}^m ; G ) \quad compare\ with \quad \min \mathcal{L} ( \tilde{\textbf{A}} ; G )
    \end{split}
\end{equation}
\noindent We can observe that $ \mathcal{G} $ will determine how to solve this optimization problem. For adversarial transferability, we introduce dedicated metrics $\varrho_1, \varrho_2$ \cite{liang2020does}, to offer a quantitative picture.
Then, we can denote the upper bound of knowledge transferability by adversarial transferability as: 
\begin{equation}
  \| \nabla  \tilde{\textbf{A}} - \nabla (\mathbf{G} \circ \hat{\textbf{A}}^m) \|^2 \leq (1-\varrho_1 \varrho_2)L^2
  \label{eq:relation}
\end{equation}
where the composite function $\mathbf{G} \circ \hat{\textbf{A}}^m$ formally denotes the knowledge transferability on $G$ to approximate $ \tilde{\textbf{A}}$ (assuming that $ \tilde{\textbf{A}}$ is $L$-Lipschitz continuous, \ie, $\| \nabla \tilde{\textbf{A}} \|_2 \leq L $). The adversarial transferability can be measured by the angle difference ($\varrho_1$) between source target gradients and the norm difference ($\varrho_2$). The inequality suggests that if adversarial transferability is high, there exists an affine transformation with the bounded norm. It provides a theoretical underpinning for self-augmentation in Sec.~\ref{sec:ad-module} and emphasizes the necessity of Sec.~\ref{sec:ad-gap}. We bridge the distribution gap to prevent mode collapse~\cite{srivastava2017veegan} in recommendation task. \\\vspace{-0.12in}

\noindent \textbf{ User-specific Patterns Modeling Through Gradient.}  Stacked GNN architecture may lead to performance degradation due to over-smoothing \cite{li2018deeper}. In contrast, our multi-modal contrastive learning is endowed with the capacity of encoding indistinguishable embeddings. 
Theoretically, by pushing the hard negatives away from the anchor, greater gradients can be obtained~\cite{wang2021understanding}. After obtaining the gradient of the negative sample (Eq.~\ref{eq:pos-neg}), we can get function $\phi(\cdot)$ by the proportional approximation (Eq.~\ref{eq:neg}). It maps the similarity of the negative sample pair to the gradient of the negative node:
    \begin{equation}
        \label{eq:prop-relat}
    \begin{split}
        \phi(x) \propto \sqrt{1-(x)^2} \cdot \exp{(x / \tau)}
    \end{split}
\end{equation}
\noindent where $x$ is the input of $\phi(\cdot)$ calculated by normalized embedding of anchor node and negative sample. $\tau$ is the temperature coefficient. 
By analyzing the gradient with Appendix.Fig.~\ref{fig:gradient}, we can conclude that our contrastive learning paradigm will assign larger gradients to hard negative samples (other users) to enhance representation discrimination, \ie, facilitating to model user-specific preference. 


\section{Evaluation}
\label{sec:eval}

In this section, we conduct experiments to validate the effectiveness of \model\ method by answering the following research questions:

\begin{itemize}[leftmargin=*]

\item \textbf{RQ1}: Can the proposed \model\ method achieve performance improvement over various state-of-the-art (SOTA) baselines?

\item \textbf{RQ2}: For key learning components in our \model, what are their impacts in boosting the recommendation performance?

\item \textbf{RQ3}: How effective is \model\ in alleviating the sparsity issue?

\item \textbf{RQ4}: How is training efficiency of \model\ as epochs increases?

\item \textbf{RQ5}: How does the hyperparameter settings impact the results?

\end{itemize}

\begin{table}[]
\caption{\textbf{Statistics of experimented datasets with multi-modal item Visual(V), Acoustic(A), Textual(T) contents.}}
\vspace{-0.15in}
\label{tab:dataset}
\small
\setlength{\tabcolsep}{0.3mm}
\begin{tabular}{cccccccccccccc}
\hline
\multirow{2}{*}{ Dataset}                   &     & \multicolumn{5}{c}{Amazon} &  & \multicolumn{3}{c}{\multirow{2}{*}{Tiktok}} &  & \multicolumn{2}{c}{\multirow{2}{*}{Allrecipes}} \\ 
   \cline{3-7} 
    &    & \multicolumn{2}{c}{Sports}  &    & \multicolumn{2}{c}{Baby}        &   &  &  &      \\ \hline \hline 
Modality  &    & V & T &  & V & T & & V & A & T & & V & T  \\   
                           Embed Dim &    & 4096 & 1024  & & 4096 & 1024 &  & 128 & 128 & 768 & & 2048 & 20     \\ \cline{1-1} \cline{3-4} \cline{6-7} \cline{9-11} \cline{13-14}  
User  &    & \multicolumn{2}{c}{35598} & & \multicolumn{2}{c}{19445}  &  & \multicolumn{3}{c}{9319}  & & \multicolumn{2}{c}{19805}  \\ 
Item   &    & \multicolumn{2}{c}{18357} & & \multicolumn{2}{c}{7050}  &  & \multicolumn{3}{c}{6710}  & & \multicolumn{2}{c}{10067}   \\  
Interactions   &    & \multicolumn{2}{c}{256308} & & \multicolumn{2}{c}{139110}  &  & \multicolumn{3}{c}{59541} & & \multicolumn{2}{c}{58922}   \\ \hline 
Sparsity   &    & \multicolumn{2}{c}{99.961\%} & & \multicolumn{2}{c}{99.899\%}  &  & \multicolumn{3}{c}{99.904\%} & & \multicolumn{2}{c}{99.970\%}   \\ \hline 
\end{tabular}
\vspace{-0.2in}
\end{table}

\begin{table*}[t]
\caption{\textbf{Performance comparison of baselines on different datasets in terms of \emph{Recall}@20, \emph{Precision}@20 and \emph{NDCG}@20.}}
\vspace{-0.15in}
\small
\label{tab:overall_performance}
\setlength{\tabcolsep}{1.55mm}
\begin{tabular}{cccclccclccclccc}
\hline
\multirow{2}{*}{Baseline} & \multicolumn{3}{c}{Tiktok}  &           & \multicolumn{3}{c}{Amazon-Baby}   &           & \multicolumn{3}{c}{Amazon-Sports} &           & \multicolumn{3}{c}{Allrecipes}                                   \\ \cline{2-4} \cline{6-8} \cline{10-12} \cline{14-16}
                          & R@20    & P@20    & N@20    &           & R@20   & P@20    & N@20    &           & R@20            & P@20            & N@20            &  & R@20               & P@20                & N@20                \\ \hline
MF-BPR                    & 0.0346  & 0.0017  & 0.0130  &           & 0.0440  & 0.0024  & 0.0200  &          & 0.0430          & 0.0023          & 0.0202          &  & 0.0137             & 0.0007              & 0.0053                    \\
NGCF                      & 0.0604  & 0.0030  & 0.0238  &           & 0.0591  & 0.0032  & 0.0261  &          & 0.0695          & 0.0037          & 0.0318          &  & 0.0165          & 0.0008          & 0.0059                     \\
LightGCN                  & 0.0653  & 0.0033 & 0.0282  &           & 0.0698  & 0.0037  & 0.0319  &          & 0.0782          & 0.0042          & 0.0369          &  & 0.0212          & 0.0010          & 0.0076                    \\ 
\hline
SGL                       &0.0603   &0.0030   &0.0238   &           &0.0678   &0.0036   &0.0296   &         &0.0779           &0.0041           &0.0361           & &0.0191           &0.0010           &0.0069                      \\
NCL                       &0.0658   &0.0034   & 0.0269 &            &0.0703   &0.0038   &0.0311   &          &0.0765           &0.0040           &0.0349           & &0.0224          &0.0010           &0.0077                      \\
HCCF                      & 0.0662  & 0.0029 & 0.0267 &           &0.0705  & 0.0037  & 0.0308  &           & 0.0779          & 0.0041          & 0.0361          & & 0.0225          & 0.0011          & 0.0082                     \\
\hline
VBPR                      & 0.0380  & 0.0018 & 0.0134  &           & 0.0486  & 0.0026  & 0.0213  &          & 0.0582          & 0.0031          & 0.0265          & & 0.0159          & 0.0008          & 0.0056                  \\
LightGCN-$M$                 & 0.0679  & 0.0034 & 0.0273  &          & 0.0726  & 0.0038  & 0.0329  &            & 0.0705          & 0.0035          & 0.0324          & & 0.0235          & 0.0011          & 0.0081                     \\
MMGCN                     & 0.0730 & 0.0036 & 0.0307 &              & 0.0640  & 0.0032  & 0.0284  &          & 0.0638          & 0.0034          & 0.0279          & & 0.0261          & 0.0013          & 0.0101                    \\
GRCN                      & 0.0804 & 0.0036 & 0.0350 &              & 0.0754  & 0.0040  & 0.0336  &          & 0.0833          & 0.0044  & 0.0377  &           & 0.0299          & 0.0015          & 0.0110  \\

LATTICE                   & 0.0843   & 0.0042    & \underline{0.0367}  & & \underline{0.0829}  & \underline{0.0044}  & \underline{0.0368}  &           & \underline{0.0915} & \underline{0.0048}  & \underline{0.0424}  &           & 0.0268          & 0.0014          & 0.0103 \\ 
CLCRec                    &0.0621  &0.0032  &0.0264 &               &0.0610  & 0.0032  & 0.0284  &           & 0.0651          & 0.0035          & 0.0301          & & 0.0231          & 0.0010          & 0.0093           \\
MMGCL                    & 0.0799 & 0.0037 & 0.0326 &               & 0.0758  & 0.0041  & 0.0331 &           & 0.0875          & 0.0046          & 0.0409          & &0.0272           & 0.0014          &    0.0102        \\
SLMRec           & \underline{0.0845} & \underline{0.0042} & 0.0353 &             & 0.0765  & 0.0043  & 0.0325  &           & 0.0829          & 0.0043           & 0.0376          & & \underline{0.0317}          & \underline{0.0016}       &    \underline{0.0118}        \\
\hline
\model\                      & \textbf{0.0921} & \textbf{0.0046}  & \textbf{0.0392}  &  & \textbf{0.0962}  & \textbf{0.0051}  & \textbf{0.0422}  &           & \textbf{0.0998} & \textbf{0.0052} & \textbf{0.0470}  &           & \textbf{0.0367} & \textbf{0.0018} & \textbf{0.0135}  \\
\emph{p}-value                   & \small{1.28$e^{-5}$} & \small{7.12$e^{-6}$} & \small{6.55$e^{-6}$} & \textbf{} & \small{2.23$e^{-6}$} & \small{7.69$e^{-6}$} & \small{8.65$e^{-7}$} &  \textbf{} & \small{7.75$e^{-6}$}         & \small{6.48$e^{-6}$}         & \small{6.78$e^{-7}$}         & & \small{3.94$e^{-4}$}             & \small{5.06$e^{-6}$}              & \small{4.31$e^{-5}$}              \\ 
Improv.                   & 8.99\%  & 9.52\%  & 6.81\% &  & 16.04\% & 15.91\% & 14.67\% & \textbf{} & 9.07\% & 8.33\% & 10.85\% & \textbf{} & 15.77\%         & 12.50\%         & 14.40\%                      \\ \hline
\end{tabular}
\vspace{-0.8em}
\end{table*}

\subsection{Experimental Settings}
\subsubsection{\bf Dataset}
Experiments are conducted on four publicly available multi-modal recommendation datasets, \ie, Tiktok, Amazon-Baby, Amazon-Sports, and Allrecipes. Data statistics with multi-modal feature embedding dimensionality are reported in Table~\ref{tab:dataset}.

\begin{itemize}[leftmargin=*]
\item \textbf{TikTok}. This data is collected from TikTok platform to log the viewed short-videos of users. The multi-modal characteristics are visual, acoustic, and title textual features of videos. The textual embeddings are encoded with Sentence-Bert~\cite{reimers2019sentence}.

\item \textbf{Amazon}. We adopt two benchmark datasets from Amazon with two item categories Amazon-Baby and Amazon-Sports~\cite{mcauley2015image}. In those datasets, textual feature embeddings are also generated via Sentence-Bert based on the extracted text from product title, description, brand and categorical information. The product images are used to generate $4096$-d visual feature embeddings of items.

\item \textbf{Allrecipes}. This dataset comes from one of the largest food-oriented social network platform by including 52,821 recipes in 27 different categories. For each recipe, its image and ingredients are considered as the visual and textual features. Following the setting in~\cite{gao2019hierarchical}, 20 ingredients are sampled for each recipe.
\end{itemize}

\vspace{-0.05in}
\subsubsection{\bf Evaluation Protocols}
To evaluate the accuracy of top-$K$ recommendation results, we adopt three widely used metrics: Recall@K (R@K), Precision@K (P@K), and Normalized Discounted Cumulative Gain (N@K). Following the settings in~\cite{wei2021hierarchical,wei2020graph}, we all-rank item evaluation strategy is used to measure the accuracy. The average scores over all users in the test set are reported.

\vspace{-0.05in}
\subsubsection{\bf Baselines}
We compare \model\ with SOTA multi-modal recommender systems, popular GNN-based collaborative filtering models, recently proposed SSL-based recommendation solutions.

\noindent i) \textbf{GNN-based Collaborative Filtering Models}
\begin{itemize}[leftmargin=*]
\item \textbf{NGCF}~\cite{wang2019neural}: The method leverages graph convolutional network to inject high-order collaborative signals into representations.
\item \textbf{LightGCN}~\cite{he2020lightgcn}: By removing the redundant transformation and non-linear activation, LightGCN simplifies the message passing for graph neural network-based recommendation.  
\end{itemize}

\noindent ii) \textbf{SSL-based Recommendation Solutions}
\begin{itemize}[leftmargin=*]
\item \textbf{SGL}~\cite{wu2021self}: This model improves the graph collaborative filtering with the incorporated contrastive learning signals using different data augmentation operators, \eg, randomly node/edge dropout and random walk, to construct contrastive representation views.
\item \textbf{NCL}~\cite{lin2022improving}: In this approach, constrastive views are generated by identifying semantic and structural neighboring nodes with EM-based clustering, to generate the positive contrastive pairs.
\item \textbf{HCCF}~\cite{xia2022hypergraph}: To supplement main recommendation objective with SSL task, HCCF leverages the hypergraph neural encoder to inject the global collaborative relations into the recommender. 
\end{itemize}

\noindent iii) \textbf{Multi-Modal Recommender Systems}
  \vspace{-0.5 em}
\begin{itemize}[leftmargin=*]
\item \textbf{VBPR}~\cite{he2016vbpr}: This is a representative work to incorporate multimedia features into the matrix decomposition framework.
\item \textbf{LightGCN-$M$}: This baseline is generated by using SOTA GNN-based CF model LightGCN as backbone and incorporate multi-modal item features during the graph-based message passing.
\item \textbf{MMGCN}~\cite{wei2019mmgcn}: It leverages graph convolutional network to propagate the modality-specific embedding and capture the modality-related user preference for micro-video recommendation.
\item \textbf{GRCN}~\cite{wei2020graph}: It is a structure-refined GCN multimedia recommender system, which yields the refined interactions to identify false-positive feedback and eliminate noisy with pruning edges.
\item \textbf{LATTICE}~\cite{zhang2021mining}: It proposes to identify latent item-item relations with the generated item homogeneous graph. Connections will be added among items with similar modality features.
\item \textbf{CLCRec}~\cite{wei2021contrastive}: This model addresses the item cold-start issue by enhancing item embeddings with multi-modal features using mutual information-based contrastive learning.
\item \textbf{MMGCL}~\cite{yi2022multi}: It incorporates the graph contrastive learning into recommender through modality edge dropout and masking.
\item \textbf{SLMRec}~\cite{tao2022self}: This method designs data augmentation on multi-modal content with two components, \ie, noise perturbation over features and multi-modal pattern uncovering augmentation. 
\end{itemize}

\subsubsection{\bf Hyperparameter Settings}
Our \model\ model is implemented with Pytorch. We adopt AdamW\cite{loshchilov2017decoupled} and Adam\cite{kingma2014adam} as the optimizer for generator and discriminator with the learning rate search range of \{$4.5e^{-4}$, $5e^{-4}$, $5.4e^{-3}$, $5.6e^{-3}$\} and \{$2.5e^{-4}$, $3e^{-4}$, $3.5e^{-4}$\}, respectively. The decay of $L_2$ regularization term is searched in \{$1.2e^{-2}$, $1.4e^{-2}$, $1.6e^{-2}$ \}. For fair comparison, the embedding dimensionality of our \model\ and other compared methods is set as 64. For our \model\ model, the number of propagation layers over graph structure is tuned from \{1, 2, 3, 4\}. 



 \vspace{-1em}

\subsection{Performance Comparison (RQ1)}
Evaluation results are reported in Table~\ref{tab:overall_performance}, in which the performance of our \model\ and the best-performed baseline are highlighted with bold and underlined, respectively. Key observations are as follows:

\begin{itemize}[leftmargin=*]
\item \model\ shows promising performance by consistently outperforming all baselines on different datasets, which demonstrates the effectiveness of our proposed new model. We attribute the performance improvement to the integrated generative and contrastive SSL components for modality-aware data augmentation. Most of recently proposed multimedia recommendation methods perform better than graph-based collaborative filtering models, which indicates the effectiveness of incorporating multi-modal context in learning modality-aware collaborative relationships. \\\vspace{-0.12in}

\item While recent studies (\ie, SGL, NCL, HCCF) attempt to augment user-item interaction modeling in a contrastive fashion, only marginal performance gains are achieved by them compared with NGCF and LightGCN. We postulate the reason is the ignorance of multi-modal contextual information for generating self-supervision signals. With the guidance of multi-modal patterns (\eg, modality-specific user preference, cross-modal relatedness), our \model\ distills modality-aware self-supervised signals to supplement the supervised task of multimedia recommendation.\\\vspace{-0.12in}

\item In comparison with multimedia recommender systems, the obvious performance improvement of our \model\ can be observed. As to state-of-the-art SSL-based methods (CLCRec, MMGCL, SLMRec), our framework \model\ leverages the generative adversarial SSL to construct modality-speicific user-item relation and contrastive cross-modal relatedness learning, for effective multi-modal augmentation. However, directly masking the modality user/item features in MMGCL may cause the important information loss, which makes the sparsity issue of inactive users even worse. Furthermore, SLMRec generates the augmented views via the pre-defined hierarchical correlations among different modality data, which may hinder the effects of self-supervised signals across various multimedia recommendation datasets.

\end{itemize}

\begin{table}[t]
    \vspace{-0.05in}
    \caption{Ablation study on key components of \model}
    \vspace{-0.15in}
    \centering
    \small
    \setlength{\tabcolsep}{1.9mm}
    \begin{tabular}{c|cc|cc|cc}
        \hline
        Data & \multicolumn{2}{c|}{Amazon-Baby} & \multicolumn{2}{c|}{Allrecipes} & \multicolumn{2}{c}{Tiktok}\\
        \hline
        Metrics & Recall & NDCG & Recall & NDCG & Recall & NDCG\\
        \hline
        \hline
        \emph{w/o-ASL} & 0.0907 & 0.0396 & 0.0326 & 0.0124  & 0.0801 & 0.0358 \\
        \emph{w/o-CL}  & 0.0924 & 0.0408 & 0.0328 & 0.0130 & 0.0821 & 0.0351 \\
        \emph{w/o-GT}  & 0.0929 & 0.0405 & 0.0325 & 0.0121 & 0.0815 & 0.0353 \\
        \emph{r/p-GAE}  & 0.0931 & 0.0411 & 0.0331 & 0.0126 & 0.0843 & 0.0364 \\
        \hline
        \emph{\model} & \textbf{0.0962} & \textbf{0.0422} & \textbf{0.0367} & \textbf{0.0135} & \textbf{0.0921} & \textbf{0.0392} \\
        \hline
    \end{tabular}
    \label{tab:module_ablation}
    \vspace{-0.1in}
\end{table}

\begin{table}[]
\caption{\textbf{Performance comparison \wrt\ different interaction sparsity degrees on Amazon-Baby and Allrecipes datasets. }}
\vspace{-0.1in}
\setlength{\tabcolsep}{0.5mm}
\small
\begin{tabular}{ccccccccc}
\hline
Baby       &  [0,4)    &  [4,6)    &  [6,9)       &  [9,13)   &  [13,100)      \\  
$^{(1e3)}$   & 9.783    & 4.081    & 2.823    & 1.649               & 1.109        \\ \hline \hline
VBPR            & 0.0166\tiny{$\uparrow$110.8\%}  &  0.0187\tiny{$\uparrow$101.6\%}  & 0.0197\tiny{$\uparrow$93.4\%}   & 0.0218\tiny{$\uparrow$87.6\%} & 0.0319\tiny{$\uparrow$115.1\%}    \\
MMGCN           & 0.0219\tiny{$\uparrow$59.8\%}  & 0.0247\tiny{$\uparrow$52.6\%}  & 0.0277\tiny{$\uparrow$37.6\%}      & 0.0315\tiny{$\uparrow$29.8\%} & 0.0608\tiny{$\uparrow$12.8\%}   \\
LATTICE         & 0.0263\tiny{$\uparrow$33.1\%}  & 0.0300\tiny{$\uparrow$25.7\%}  & 0.0337\tiny{$\uparrow$13.1\%}    & 0.0366\tiny{$\uparrow$11.8\%} & 0.0619\tiny{$\uparrow$10.8\%}    \\ 
SLMRec         & 0.0269\tiny{$\uparrow$30.1\%}  & 0.0291\tiny{$\uparrow$29.6\%}  & 0.0302\tiny{$\uparrow$26.2\%}    & 0.0318\tiny{$\uparrow$28.6\%} & 0.0611\tiny{$\uparrow$12.3\%}    \\ 
Ours            & 0.0350  & 0.0377  & 0.0381  & 0.0409 & 0.0686   \\ \hline \hline 
Allrecipes       &  [0,2)    &  [2,3)    &  [3,4)      &  [4,5)   &  [5,10)    \\  
$^{(1e3)}$ & 8.720    & 4.643    & 3.221    & 1.790               & 1.431     \\ \hline \hline
VBPR               & 0.0039\tiny{$\uparrow$164.1\%}  &  0.0051\tiny{$\uparrow$132.7\%}  & 0.0056\tiny{$\uparrow$141.1\%}   & 0.0057\tiny{$\uparrow$127.6\%} & 0.0069\tiny{$\uparrow$175.4\%}   \\
MMGCN               & 0.0067\tiny{$\uparrow$53.7\%}  & 0.0084\tiny{$\uparrow$44.1\%}  & 0.0100\tiny{$\uparrow$35.0\%}      & 0.0104\tiny{$\uparrow$26.9\%} & 0.0134\tiny{$\uparrow$41.8\%}   \\
LATTICE            & 0.0072\tiny{$\uparrow$43.1\%}  & 0.0088\tiny{$\uparrow$37.5\%}  & 0.0105\tiny{$\uparrow$28.6\%}    & 0.0099\tiny{$\uparrow$33.3\%} & 0.0159\tiny{$\uparrow$19.5\%}   \\ 
SLMRec         & 0.0085\tiny{$\uparrow$21.2\%}  & 0.0099\tiny{$\uparrow$22.2\%}  & 0.0116\tiny{$\uparrow$16.4\%}    & 0.0114\tiny{$\uparrow$15.8\%} & 0.0174\tiny{$\uparrow$9.2\%}    \\ 
Ours            & 0.0103  & 0.0121  & 0.0135  & 0.0132 & 0.0190   \\ \hline
\end{tabular}
\label{-0.1in}
\label{tab:sparsity}
\end{table}

\vspace{-0.1in}
\subsection{In-Depth Analysis (RQ2 and RQ3)}

\subsubsection{\bf Ablation Study (RQ2)}
\label{sec:ablation}
Experiments are conducted to justify the importance of key components in \model\ with the details. 

\noindent (1) We first disable the adversarial generative self-augmentation in the variant w/o-ASL. As shown in Table~\ref{tab:module_ablation}, the performance of w/o-ASL without our adversarial SSL decreases sharply compared with our \model, demonstrating the strength of our designed generative augmentation with modality-guided self-supervised learning. \\\vspace{-0.12in}

\noindent (2) We ablate the cross-modal contrastive learning paradigm with the variant w/o-CL. The superior model accuracy further reveals the significance of our contrastive augmentation by exploring the modality-wise dependencies for multimedia recommendation. \\\vspace{-0.12in}

\noindent (3) To emphasize the rationality of our adversarial SSL method, we make another comparison between \model\ and another variant (w/o-GT) without the Gumbel-based transformation. The observed performance gain reflects the improvements of our designed Gumbel-based transformation in enhancing the adversarial learning in addressing the distribution gap issue. \\\vspace{-0.12in}

\noindent (4) Compared with r/p-GAE, which replaces our adversarial learning component with graph autoencoder 
for generating multi-modal user-item relational patterns, our \model\ always performs the best. This indicates the superiority of capturing implicit user-item relations with the awareness of modality-aware preference using our framework for useful data augmentation in generative fashion.

\subsubsection{\bf Evaluation \wrt\ Data Sparsity (RQ3)}
To examine the effectiveness of \model\ in addressing the sparsity issue, we separately evaluate the recommendation accuracy with different interaction frequencies of users. According to the presented results (measured by NDCG@20) in Table~\ref{tab:sparsity}, it is obvious that our \model\ consistently outperforms the compared approaches under different interaction sparsity degrees, which further validates the rationality of our new self-supervised learning in mitigating the data sparsity issue to some degree. With the self-augmented multi-modal signals, our \model\ can generate more accurate representations via effectively transferring multi-modal knowledge in an expressive manner. 

\subsubsection{\bf Model Convergence Analysis (RQ4)}
\begin{figure}[t]
	\centering
    \includegraphics[width=0.9998\columnwidth]{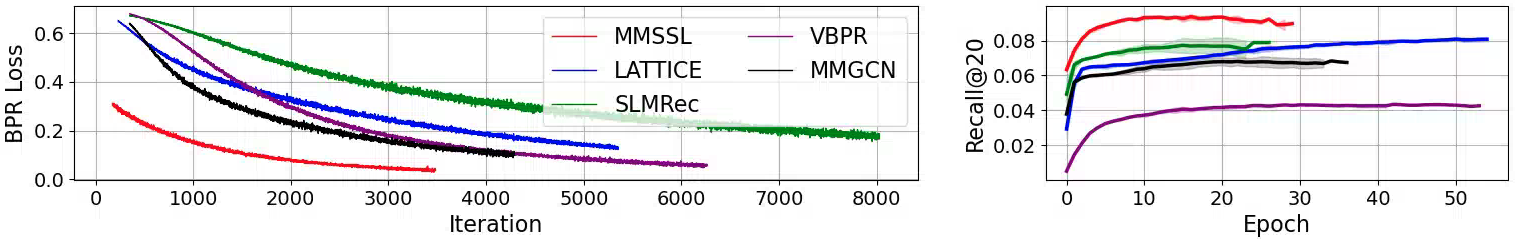}
	\vspace{-0.3in}
	\caption{Training curves of \model\ and compared methods.}
        \Description[Training curves of \model\ and compared methods.]{In Figure~\ref{fig:convergence}, we show the training curves of \model\ and compared methods on Amazon-Baby dataset, as the number of iterations and epochs increases. From the results, the faster converge speed of our \model\ method is obviously observed, which suggests the advantage of \model\ in training efficiency, meanwhile maintaining superior recommendation accuracy. This indicates that our incorporated multi-modal self-supervision signals bring positive effects to learn useful gradients during model optimization phase for fast convergence.}
	\label{fig:convergence}
	\vspace{-0.15in}
\end{figure}

In this section, we study the impact of our multi-modal self-supervised learning paradigm in model training efficiency with convergence analysis. In Figure~\ref{fig:convergence}, we show the training curves of \model\ and compared methods on Amazon-Baby dataset, as the number of iterations and epochs increases. From the results, the faster converge speed of our \model\ method is obviously observed, which suggests the advantage of \model\ in training efficiency, meanwhile maintaining superior recommendation accuracy. This indicates that our incorporated multi-modal self-supervision signals bring positive effects to learn useful gradients during model optimization phase for fast convergence. The parameter inference procedures of compared learning methods (\eg, LATTICE, MMGCN) require sufficient training labels (\ie, user interactions) to learn good gradients, which are limited by the label shortage issue in practical scenarios.

\subsection{Impact Study of Hyperparameters (RQ4)}

\begin{figure}[t]
	\centering
y    \includegraphics[width=0.9\columnwidth]{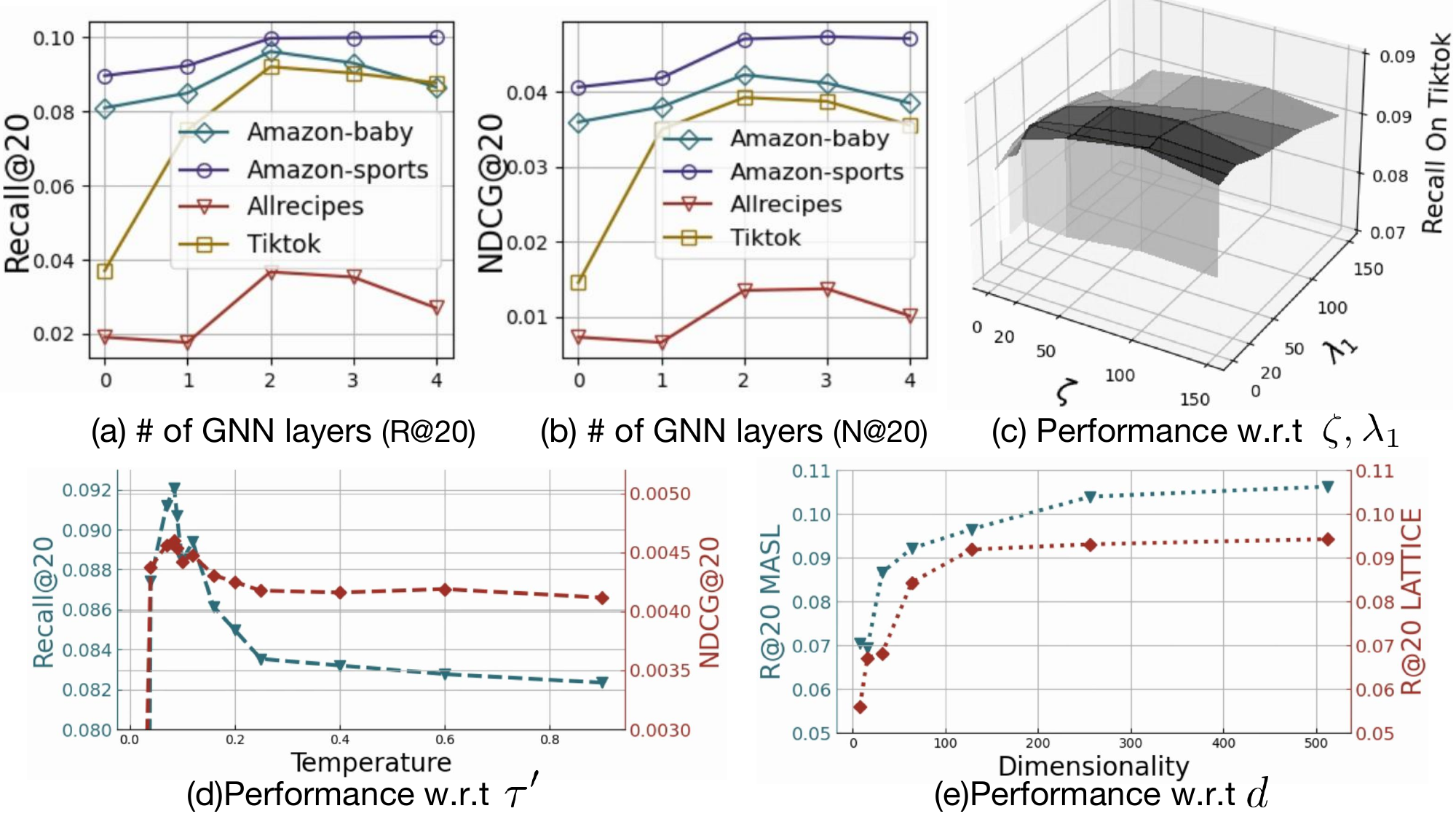}
	\caption{Impact study of hyperparameters in \model.}
         \Description[Impact study of hyperparameters in \model.]{We examine the sensitivity of several important parameters of our MASL model on different datasets, including \# of GNN layers $L$, augmentation factors $\zeta$, $\lambda_1$ in Adversarial SSL, temperature parameter $\tau$ and latent dimensionality $d$.}
	\label{fig:parameter}
	\vspace{-0.1in}
\end{figure}

In this section, we examine the sensitivity of several important parameters of our \model\ model on different datasets.

\begin{itemize}[leftmargin=*]
    \item \textbf{Effect of \# of GNN layers $L$}. We first investigate the influence of GNN model depth by varying the number of message passing layers $L$ from 1 to 4. As we can see in Figure~\ref{fig:parameter}, our method performs the best with 2 or 3 graph layers. As the model goes deeper, the oversmoothing issue raises in the encoded embeddings, and thus decreasing the recommendation performance.
    
    \item \textbf{Effect of augmentation factors $\zeta$, $\lambda_1$, in Adversarial SSL}. As presented in Section 3.1.3 and 3.1.4, $\zeta$, $\lambda_1$ are augmentation factors in addressing the distribution gap in our multi-modal adversarial self-augmentation paradigm. Following similar settings in~\cite{petzka2017regularization}, $\zeta$ is selected from the range of [0, 10, 20, 50, 100, 150]. The best performance is obtained with $\zeta=100$. $\lambda_1$ is tuned from [0, 1, 10, 20, 50, 100, 150] and performs the best when $\lambda_1$=1. The evaluation results indicate that the augmentation factors with appropriate values are effective to alleviate the distribution gap in our minmax optimization for adversarial self-augmentation.
    
    \item \textbf{Effect of temperature parameter $\tau$}. We tune the hyperparameter $\tau$ from ($0,0.9$) to control the agreement strength between the instances of positive pairs. From the results on TikTok dataset, the best performance is achieved with $\tau=0.085$. 
    
    \item \textbf{Effect of latent dimensionality $d$}. The embedding dimensionality $d$ of our model is searched from ($2^3$, $2^4$, $2^5$, $2^6$, $2^7$, $2^8$, $2^9$). Comparing the performance of our \model\ and the best-performed baseline (LATTICE), we observe the consistent performance improvement achieved by our method, which further justifies the effectiveness of our modality-aware self-supervision for enhancing model robustness by learning augmented representations.
\end{itemize}


\section{Related Work}
\label{sec:relate}

\noindent \textbf{GNN-based Recommender Systems}. 
Graph neural networks have been widely adopted in recommender systems to model various relationships in different recommendation scenarios. For example, 1) user-item interactions in collaborative filtering (\eg, LightGCN~\cite{he2020lightgcn}, LR-GCCF~\cite{chen2020revisiting}); 2) user connections in social recommendation (\eg, GraphRec~\cite{fan2019graph}, KCGN~\cite{huang2021knowledge}); 3) item-item temporal transitional relationships in sequential recommendation (\eg, GCE-GNN~\cite{wang2020global}, SURGE~\cite{chang2021sequential}); and 4) entity-item dependencies in knowledge graph-enhanced recommender (\eg, KGAT~\cite{wang2019kgat}). Motivated by these research studies, our \model\ method adopts GNN as the backbone to model the high-order collaborative relationships with the injection of multi-modal contextual information. \\\vspace{-0.1in}

\noindent \textbf{Self-Supervised Learning for Recommendation}.
Recently, self-supervised learning (SSL) has shown its effectiveness in addressing label scarcity for recommendation~\cite{lin2022improving,mengru2023contrastive}. At its core is to augment original supervision signals with the incorporated auxiliary learning task. For graph augmentation with contrastive learning, NCL~\cite{lin2022improving}, CML~\cite{wei2022contrastive} and HCCF~\cite{xia2022hypergraph} propose to generate SSL signals via contrasting positive node pairs based on various augmentation operators, \eg, random walk graph sampling and semantic neighbor identification. For SSL-based sequence augmentation, CL4SRec~\cite{xie2022contrastive} augments item sequence in three different ways, \ie, crop, mask and reorder. S$^3$-Rec~\cite{zhou2020s3} performs contrastive learning among item sequence and attribute sequence. Additionally, for augmenting relational learning in social recommendation, MHCN~\cite{yu2021self} designs SSL task to capture high-order connectivity using mutual information maximization. Different from these works, for robust multi-modal user preference learning, we creatively design a new SSL recommender which adversarially trains a modality-aware neural graph generator to integrate with cross-modal contrastive augmentation.\\\vspace{-0.1in}

\noindent \textbf{Multimedia Recommendation}. Many efforts have been devoted to enhancing recommender systems by incorporating multimedia content. One representative early study VBPR~\cite{he2016vbpr} extends matrix factorization to integrate both id-corresponding embeddings and multimedia feature embeddings of items. To improve the user-item relation modeling with multimedia content, attention mechanisms are used in ACF~\cite{chen2017attentive} and VECF~\cite{chen2019personalized} 
to capture complex user preference. In recent years, graph neural networks have been demonstrated as powerful solutions for multimedia recommendation by capturing high-order dependent structures among users and items. For example, graph convolutional network used in MMGCN~\cite{wei2019mmgcn} and GRCN~\cite{wei2020graph}, and graph attention mechanism applied in MKGAT~\cite{sun2020multi}. In this work, we propose a new multimedia recommender system which addresses the limitation of heavily rely on abundant labels in existing methods with a dual-stage SSL paradigm.

\section{Conclusion}
\label{sec:conclusoin}

In this work, we tackle the problem of multimedia recommendation by proposing a multi-modal self-supervised learning model \model. In \model, we design a novel SSL task with modality-aware adversarial perturbation to capture multi-modal user preference under sparse interaction labels. In addition, we introduce the cross-modal contrastive learning paradigm to enable the dependency modeling across different modality-specific user interaction patterns. Through extensive experiments on several public datasets, the proposed \model\ with effective self-supervision can achieve significant performance gains compared with various baselines. Future studies include extension of \model\ to encode multi-interest preference of users with diverse latent embeddings. To incorporate multi-dimensional interests into our encoded user embeddings, the multi-modality information can be fed into our interest identification component by clustering multimedia items. In addition, in our future work, it is interesting to enhance the explainablity of our \model\ by developing a GNN-based explainer to learn causal effects on modality-aware user-item interaction graph. It will facilitate the understanding of user preference with intuitive explanations.




\clearpage

\clearpage
\balance
\bibliographystyle{ACM-Reference-Format}
\bibliography{sample-base}
\clearpage
\appendix \section{Appendix}
\balance
\label{sec:appendix}

 In this section, in-depth details are included to: i) analyze the user preference diversity brought by contrastive learning paradigm from the perspective of gradients;
ii) discuss the theoretical basis of self-augmented adversarial collaborative knowledge transfer; iii) analyze the time complexity of our \model; iv) further justify the effectiveness of our proposed \model\ with additional experiments. 

\subsection{\bf Derivation of Negative Sample Gradient}
Inspired by works \cite{wang2021understanding, wu2021self} which state contrastive loss with hardness-aware ability can push away the hard negative instances from the anchor by giving greater gradients, we leverage this property to tackle the over-smoothing issue~\cite{li2018deeper} when encoding high-order collaborative signals and learning distinguishable embeddings.

\noindent \textbf{Gradient of Positive and Negative Samples.}We provide the gradient derivation procedure to support the point described in Sec.~\ref{sec:cl-theory}.
The partial derivative of the anchor point is computed using $\mathcal{L}_{ \text{CL}_u}$ to examine the impact of samples on gradients:
\begin{equation}
    \label{eq:pos-neg}  
    \begin{split}
     & \frac{\partial{ \mathcal{L}_{ \text{CL}_u} }}{\partial{\textbf{q}_u}}  =
     \frac{\partial{}}{\partial{\textbf{q}_u}} \left( 
     -log \frac{\exp{(\textbf{q}_u \cdot \textbf{q}_P / \tau)}}{ \sum_{\mathcal{U}_A}{\exp{(\textbf{q}_u \cdot \textbf{q}_A / \tau)}}}
     \right)\\\\ \vspace{5ex}
     & = \frac{\partial{}}{\partial{\textbf{q}_u}}\left(-log \exp{(\textbf{q}_u\cdot \textbf{q}_P / \tau)} \right)  +   \frac{\partial{}}{\partial{\textbf{q}_u}}\left(log \sum_{\mathcal{U}_A}{\exp{(\textbf{q}_u \cdot \textbf{q}_A / \tau)} } \right) \\\\
     & = -\frac{1}{\tau} \cdot \textbf{q}_P + \frac{1}{\tau} \frac{\sum_{\mathcal{U}_A}{\textbf{q}_A \cdot \exp{(\textbf{q}_u \cdot \textbf{q}_A / \tau)}}}{\sum_{\mathcal{U}_A}{\exp{(\textbf{q}_u \cdot \textbf{q}_A / \tau)}}}\\\\  \vspace{5ex}
     & = \frac{1}{\tau} \left( 
     \frac{\sum_{\mathcal{U}_N}{\textbf{q}_N \cdot \exp{(\textbf{q}_u \cdot \textbf{q}_N / \tau)} }}{\sum_{\mathcal{U}_A}{\exp{(\textbf{q}_u \cdot \textbf{q}_A / \tau)}}}
     +\frac{{\textbf{q}_P \cdot \exp{(\textbf{q}_u \cdot \textbf{q}_P / \tau)} }}{\sum_{\mathcal{U}_A}{\exp{(\textbf{q}_u \cdot \textbf{q}_A / \tau)}}}
     - \textbf{q}_P     
     \right) \\\\ \vspace{2.5ex}
    & = \begin{matrix} \underbrace{   \frac{1}{\tau \cdot \parallel \textbf{e}_u \parallel} \cdot
    \frac{\sum_{\mathcal{U}_N}{ \left(\textbf{q}_N-(\textbf{q}_u \cdot \textbf{q}_N) \cdot \textbf{q}_u \right) \cdot \exp{(\textbf{q}_u \cdot \textbf{q}_N / \tau)} }}{\sum_{\mathcal{U}_A}{\exp{(\textbf{q}_u \cdot \textbf{q}_A / \tau)}}}   } \\ negative \end{matrix} \\  \vspace{1.5ex}
    & + \begin{matrix} \underbrace{  \frac{1}{\tau \cdot \parallel \textbf{e}_u \parallel} \cdot   \left( \textbf{q}_P-(\textbf{q}_u \cdot \textbf{q}_P) \cdot \textbf{q}_u \right) \cdot \left(\frac{\exp{(\textbf{q}_u \cdot \textbf{q}_P / \tau)}}{\sum_{\mathcal{U}_A}{\exp{(\textbf{q}_u \cdot \textbf{q}_A / \tau)}}} - 1 \right)    } \\ positive\end{matrix}   \\
    \end{split}
\end{equation}
where $\textbf{q}_u$ represents the normalized (\eg, $\textbf{q}_u = \textbf{e}_u/\parallel \textbf{q}_u \parallel$) anchor node mapped into the same hyperspace with other nodes. $\textbf{q}_P, \textbf{q}_N, \textbf{q}_A \in\mathbb{R}^{d\times 1}$ are the instances from the positive, negative, and the entire set, respectively. The chain rule employed in Eq.~\ref{eq:pos-neg} is detailed in Eq.~\ref{eq:norm1} and Eq.~\ref{eq:norm2}.
Consequently, it can be clearly concluded that the gradient of user $u$ can be determined using both the positive and negative pairs after the calculation in Eq.~\ref{eq:pos-neg}.

\begin{figure}[H]
    \centering
    \includegraphics[width=0.98\columnwidth]{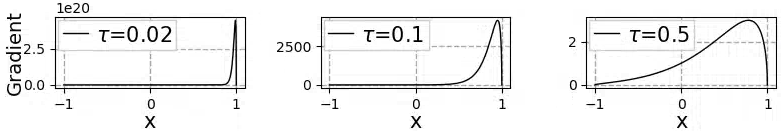}
    \vspace{-0.1in}
    \caption{Gradient function $\phi(x)$ in Eq.~\ref{eq:prop-relat} when $\tau$ = 0.02, $\tau$ = 0.1 and $\tau$ = 0.5. $x$ is the similarity between positive and
negative instances. This demonstrates that the gradient increases with decreasing temperature coefficient $\tau$.}
    \Description[Gradient function $\phi(x)$.]{Gradient function $\phi(x)$ in Eq.~\ref{eq:prop-relat} when $\tau$ = 0.02, $\tau$ = 0.1 and $\tau$ = 0.5. $x$ is the similarity between positive and negative instances. This demonstrates that the gradient increases with decreasing temperature coefficient $\tau$.}
\label{fig:gradient}
\end{figure}

\noindent \textbf{Proportional Relationship of Negative Sample Gradient.} To get $\phi(x)$ which maps the similarity of the negative sample pair to the gradient of negative node, we focus on the gradient produced by negative cases shown as follows:
\begin{equation}
    \begin{split}
    \frac{ \left(\textbf{q}_N-(\textbf{q}_u \cdot \textbf{q}_N)\cdot \textbf{q}_u \right) \cdot \exp{(\textbf{q}_u \cdot \textbf{q}_N / \tau)} }{\sum_{\mathcal{U}_A}{\exp{(\textbf{q}_u \cdot \textbf{q}_A / \tau)}}} \\
    \end{split}
\end{equation}
The norm of the corresponding gradient is proportional to the term:
\vspace{-1.0em}

\begin{equation}
        \label{eq:neg}
    \begin{split}
\parallel & \textbf{q}_N-(\textbf{q}_u \cdot \textbf{q}_N)\cdot \textbf{q}_u  \parallel  \left| \frac{  \exp{(\textbf{q}_u \cdot \textbf{q}_N / \tau)} }{\sum_{\mathcal{U}_A}{\exp{(\textbf{q}_u \cdot \textbf{q}_A / \tau)}}} \right| \\
& \Rightarrow \sqrt{1-(\textbf{q}_u \cdot \textbf{q}_N)^2} \left| \frac{ \exp{(\textbf{q}_u \cdot \textbf{q}_N / \tau)} }{\sum_{\mathcal{U}_A}{\exp{(\textbf{q}_u \cdot \textbf{q}_A / \tau)}}} \right|\\
& \propto \sqrt{1-(\textbf{q}_u \cdot \textbf{q}_N)^2} \cdot \exp{(\textbf{q}_u \cdot \textbf{q}_N / \tau)}
    \end{split}
\end{equation}
\noindent Given that both $ \textbf{q}_u$ and $ \textbf{q}_N$ are unit vectors, we introduce the variable $x$ with the definition of $x = q_u \cdot q_N \in [-1,1]$ to abbreviate the conclusion of Eq.~\ref{eq:neg} into function $\phi(\cdot)$:
    \begin{equation}
        \label{eq:prop-relat}
    \begin{split}
        \phi(x) \propto \sqrt{1-(x)^2} \cdot \exp{(x / \tau)}
    \end{split}
\end{equation}
To facilitate the analysis, we plot the gradient function $\phi(x)$ in Eq.~\ref{eq:prop-relat} in Fig.~\ref{fig:gradient}. We can observe that the gradient of negative samples will rise as $x$ increases. 
In other words, our contrastive learning paradigm will assign larger gradients to hard negative samples (other users) so as to enhance the discrimination of user representations.
\vspace{-1.0em}

\noindent \textbf{Chain Rule for Normalized Embedding.}
\label{sec:norm-embed}
This part further discusses Eq.\ref{eq:pos-neg}. Specifically, the gradient of the loss with respect to $\textbf{e}_u$ is related to that of $\textbf{q}_u$ via the chain rule presented as: 
\begin{equation}
\label{eq:norm1}
\frac{ \partial{\mathcal{L}_u(\textbf{q}_u)} }{ \partial{ \textbf{e}_u }} =  \frac{ \partial{\mathcal{L}_u(\textbf{q}_u)} }{ \partial{\textbf{q}_u}}
\frac{ \partial{\textbf{q}_u} }{ \partial{\textbf{q}_u} } 
\end{equation}
\begin{equation}
\label{eq:norm2}
    \begin{split}
     \frac{\partial{\textbf{q}_u}}{\partial{\textbf{e}_u}} & = \frac{\partial{}}{\partial{\textbf{e}_u}}\left( \frac{\textbf{e}_u}{\parallel \textbf{e}_u \parallel} \right) 
                 = \frac{1}{\parallel \textbf{e}_u \parallel}I - \textbf{e}_u\left(\frac{\partial{(1/\parallel \textbf{e}_u\parallel)}}{\partial{\textbf{e}_u}} \right)^T \\
                & = \frac{1}{\parallel \textbf{e}_u \parallel}\left(I - \frac{\textbf{e}_u \textbf{e}_u^T}{\parallel \textbf{e}_u \parallel^2}   \right) 
                 = \frac{1}{\parallel \textbf{e}_u \parallel}\left(I - \textbf{q}_u \textbf{q}_u^T \right)
    \end{split}
\end{equation}

\cleardoublepage
\clearpage
\clearpage

\end{document}